\documentclass[11pt]{article}
\usepackage[T1]{fontenc}
\usepackage{lmodern}
\usepackage[margin=1in]{geometry}
\usepackage[tablesonly,nolists]{endfloat}
\usepackage{setspace}
\usepackage{threeparttable}
\usepackage{amsmath,amssymb,amsthm}
\usepackage{natbib}
\usepackage{url}
\usepackage{booktabs}
\newtheorem{thm}{Theorem}

\usepackage{titlesec}
\titleformat{\section}
  {\normalfont\large\bfseries\centering}
  {\thesection.}{0.5em}{}
  
\title{Semiparametric Inference for Marginal Win Odds with Censored Time-to-Event Outcomes}
\author{Zhiguo Li\\
{\small Department of Biostatistics and Bioinformatics, Duke University}\\
{\small Email: zhiguo.li@duke.edu}}

\date{}

\begin{document}

\maketitle

\begin{abstract}
Win statistics provide clinically interpretable measures for comparing treatment groups with prioritized composite outcomes, but efficient covariate-adjustment methods for win statistics with right censored time-to-event outcomes remain limited. We develop a semiparametric framework
for marginal win probability and win odds with prioritized time-to-event outcomes, including semi-competing risks, while also accommodating scalar survival outcomes as a special case. The estimand compares independent draws from the marginal distributions of the potential outcomes  under treatment
and control, which preserves a marginal causal interpretation.
We derive the efficient influence function under conditional independent censoring, allowing censoring to depend on baseline covariates and predictable event history. The resulting cross-fitted one-step estimator incorporates covariate adjustment and censoring martingale augmentation,
allows flexible estimation of nuisance parameters, and attains the semiparametric efficiency bound under suitable  conditions on consistency and convergence rate of nuisance parameters. Simulation studies show that the estimator has small bias, accurate standard error estimation and confidence interval
coverage, and substantial efficiency gain relative to the inverse-probability-of-censoring weighted (IPCW) estimator. An application to a randomized breast cancer trial comparing tamoxifen combined with radiotherapy against tamoxifen alone demonstrates improved precision, with an approximately 25\% reduction in estimated variance at five years.
\end{abstract}

\vspace{1em}

\section{Introduction}
\label{sec1}

Composite endpoints are widely used in clinical trials to capture multiple aspects of disease progression and treatment benefit. In many applications, the components of a composite endpoint differ substantially in clinical importance. For example, in cardiovascular trials, death is typically considered more clinically important than hospitalization, and in oncology trials, death is often prioritized over disease progression or other nonterminal events. Conventional analyses of composite time-to-event endpoints, such as time to first event, often ignore this ordering and treat all component events as equivalent. This may reduce clinical interpretability when less severe events occur more frequently than more severe events.

Win statistics provide an appealing alternative for analyzing such prioritized outcomes. These methods compare outcomes between treatment groups through pairwise patient comparisons according to a prespecified
hierarchical rule. The most clinically important outcome is compared first, and outcomes with lower priority are considered only when the comparisons of the more important outcomes are tied, which may occur when neither patient experiences the corresponding event within a prespecified time horizon. A treated patient is said to ``win'' against a
control patient if the treated patient has a more favorable outcome at the first component outcome that distinguishes the pair, while the control patient wins if the reverse is true, and otherwise the pair
is classified as tied. Summary measures such as the win probability, win difference, win odds, and win ratio have become increasingly popular for analyzing prioritized composite outcomes in clinical trials \citep[e.g.,][]{Buyse2010,Pocock2012,BebuLachin2016,Brunner2021,Pocock2024}. These measures preserve the clinical hierarchy among outcome components while providing interpretable treatment comparisons.

Among these measures, we focus on the marginal win probability and the corresponding win odds. The win probability has a direct probabilistic interpretation, while the win odds is a one-to-one transformation of the same quantity. Under the convention that ties receive weight $1/2$, ties contribute equally to the two treatment groups, and the win difference is likewise determined by the win probability. In contrast, the win ratio is the ratio of win and loss probabilities and does not incorporate tied pairs into the effect measure. Consequently, it may indicate
a large relative treatment difference even when most pairwise comparisons are tied, as can occur with restricted or ordinal outcomes.

A general causal framework for marginal pairwise comparison estimands was developed by \citet{Mao2018}, who studied causal U-statistics defined through comparisons between marginal distributions of potential outcomes. This framework provides a rigorous foundation for win probability and related parameters and develops inverse-probability-weighted, locally efficient, and doubly robust estimators. Because the two potential outcomes of the same individual cannot be observed at the same time, their joint distribution is generally not identifiable without additional assumptions. The marginal formulation instead compares draws from two independent subjects, one from the treatment group and one from the control group,  which are identifiable under standard causal identification assumptions.

The development in \citet{Mao2018}, however, concerns fully observed outcomes. Many clinical trial endpoints of interest are time-to-event outcomes subject to right censoring. In such settings, the complete prioritized outcome may be
only partially observed, and pairwise comparison of the complete data between two participants may not be directly available from the observed data. This creates an additional missing data problem beyond the pairwise structure of the estimand.

In addition to accounting for censoring, improving estimation efficiency through covariate adjustment is another important consideration. Covariate adjustment is a standard strategy for improving precision of estimation  in randomized trials, particularly when baseline covariates are prognostic for the outcome
\citep{Tsiatis2008,Zhang2008,Benkeser2021,DandlHothorn2026,FDA2023}. Importantly, baseline covariates can improve estimation efficiency without changing the marginal win probability or win odds estimand.
Recent work has extended covariate adjustment and semiparametric methods to win type parameters. \citet{Scheidegger2026} connected the win odds to the marginal probabilistic index and showed how baseline covariate
adjustment can improve precision. \citet{Cao2025} developed covariate-adjusted win statistics for randomized trials with ordinal outcomes using weighting and augmented weighting estimators. For censored survival outcomes, \citet{Martinussen2025} developed debiased estimation of causal net benefit with flexible nuisance estimation based on efficient influence functions. Censoring in prioritized win statistics has also been addressed using inverse-probability-of-censoring weighting (IPCW). \citet{Dong2020IPCW} developed an IPCW-adjusted win ratio for right-censored outcomes, and \citet{Dong2021Dependent} extended this approach to dependent censoring. More recently, \citet{Huang2026} proposed doubly robust and efficient win ratio regression for prioritized outcomes, using a future-score correction that incorporates conditional expectations of unobserved pairwise score contributions.

These developments provide important foundations for efficient estimation of win type treatment effects under censoring, but they address different estimands and settings. \citet{Martinussen2025} focused on causal net benefit for scalar survival outcomes, whereas \citet{Huang2026} considered regression coefficients defined through time-integrated pairwise comparisons. The observed-data efficient influence function for the fixed-horizon marginal win probability with prioritized semi-competing risk outcomes and history-dependent censoring has not been developed in these frameworks.

In this article, we develop a semiparametric framework for marginal win probability and win odds estimands with censored
time-to-event outcomes. The estimand is defined through restricted potential outcomes independently of the censoring
mechanism. We characterize the observed-data efficient influence function and semiparametric efficiency bound for
the marginal win probability under right censoring. The influence function combines an inverse-probability-of-censoring
weighted component, a baseline covariate projection, and a censoring martingale augmentation involving predictable
event history. Based on this representation, we develop a cross-fitted one-step estimator \citep{Chernozhukov2018}
that accommodates flexible, potentially nonparametric estimation of nuisance functions and attains the
semiparametric efficiency bound under suitable conditions on the consistency and rate of convergence of nuisance function estimation. We also develop direct variance estimation based on the influence function and inference for the corresponding win odds.

Although we focus on multivariate prioritized outcomes, the proposed framework applies to both scalar time-to-event outcomes and multivariate prioritized outcomes, including semi-competing risks. The scalar right-censored case connects the proposed methodology to existing work on censored Mann--Whitney and probabilistic-index estimands \citep{KoziolJia2009}. For prioritized semi-competing risks, the observed event history provides additional information about the terminal event, nonterminal event, and censoring processes. Incorporating this information can improve precision relative to standard IPCW estimation while preserving the marginal estimand.

The remainder of the article is organized as follows. Section~2 introduces the setup and defines the marginal
win probability and win odds estimands. Section~3 establishes identification and derives the efficient influence function. Section~4 develops the proposed estimator and its asymptotic properties.
Section~5 evaluates finite-sample performance through simulation studies. Section~6 illustrates the method
using data from a randomized breast cancer trial comparing tamoxifen combined with radiotherapy against
tamoxifen alone, and then we conclude  with a discussion in Section~7.

\section{Setup and Marginal Win Odds}

We consider two outcome settings throughout: a scalar right-censored time-to-event outcome and a two-component prioritized semi-competing risk outcome. The latter is particularly relevant to prioritized
composite endpoints because a terminal event such as death, which is of higher importance, can preclude a subsequent nonterminal event, such as hospitalization or disease progression. This setting also allows us to account for
dependence between event processes and censoring that may depend on the observed event history.

Consider a study with binary treatment assignment \(A\in\{0,1\}\), and let \(X\) denote baseline covariates.
For each treatment level \(a\in\{0,1\}\), let \(Y^\tau(a)\) denote the restricted potential outcome up to a fixed time
horizon \(\tau\), with its form depending on the outcome setting. For the scalar time-to-event setting, let \(T(a)\) denote the potential event time under treatment \(a\), and define
\(
Y^\tau(a)=T(a)\wedge\tau.
\)
For the two-component prioritized semi-competing risk setting, let
\(T_1(a)\) and \(T_2(a)\) denote the potential event times of the higher priority terminal event and lower priority nonterminal event, respectively. We define
\(
T_1^\tau(a)=T_1(a)\wedge \tau,
\
T_2^\tau(a)=T_2(a)\wedge T_1(a)\wedge \tau,
\)
where the additional truncation by \(T_1(a)\) in the definition of
\(T_2^\tau(a)\) reflects the semi-competing risk structure: the lower priority
nonterminal event is only used while the subject remains free of the terminal
event. Thus, \(T_2^\tau(a)\) does not require defining a hypothetical
nonterminal event time after the terminal event. In this setting, we set
\(
Y^\tau(a)=\{T_1^\tau(a),T_2^\tau(a)\}.
\)

Let
\(
h:\mathcal Y\times\mathcal Y\rightarrow [0,1]
\)
be a prespecified comparison function, where \(\mathcal Y\) denotes the common sample space of the restricted potential
outcomes \(Y^\tau(a)\), \(a\in\{0,1\}\). For two generic restricted outcomes \(y_1,y_0\in\mathcal Y\),
the value \(h(y_1,y_0)\) measures the degree to which \(y_1\) is preferred to \(y_0\). In the marginal win probability below, \(y_1\) and \(y_0\) represent independent draws from the marginal distributions of \(Y^\tau(1)\) and \(Y^\tau(0)\), respectively, rather than the two potential outcomes of the same individual.

For the win probability, we use the scoring rule
\(
h(y_1,y_0)=I(y_1\succ y_0)+I(y_1\sim y_0)/2,
\)
where \(y_1\succ y_0\) indicates that \(y_1\) wins against \(y_0\), and
\(y_1\sim y_0\) indicates a tie. For prioritized outcomes, comparisons are
performed sequentially according to the prespecified hierarchy.
Since \(y_1\) and \(y_0\) are restricted outcomes, the time horizon
\(\tau\) is already incorporated into their components, and no
additional restriction on event times is needed in the comparison rule.
In the two-component case, for \(y_1=(t_{11},t_{12})\) and
\(y_0=(t_{01},t_{02})\), \(y_1\succ y_0\) if \(t_{11}>t_{01},\)
or if \(t_{11}=t_{01}\ \text{and}\ t_{12}>t_{02}.\)
A tie occurs if \(t_{11}=t_{01}\ \text{and}\ t_{12}=t_{02}.\)

Let \(Y_1^\tau(1)\) and \(Y_0^\tau(0)\) denote independent copies of
\(Y^\tau(1)\) and \(Y^\tau(0)\), respectively. Following the causal
U-statistic formulation of \citet{Mao2018}, define
\[
\psi_h(\tau)
=
E\left[
h\{Y_1^\tau(1),Y_0^\tau(0)\}
\right].
\]
This parameter is the marginal probability that an outcome drawn from the distribution of the potential outcome of the treatment group  is preferred to an outcome drawn from the distribution of the potential outcome of the control group, with ties counted as one half under the scoring rule above. Here \(Y_1^\tau(1)\) and \(Y_0^\tau(0)\) are written as independent generic draws to distinguish them from the pair \(\{Y^\tau(1),Y^\tau(0)\}\),
which  denotes the two potential outcomes for the same subject under the two treatment levels. Thus, \(\psi_h(\tau)\) depends only on the two distributions of the marginal potential outcomes and does not require the joint cross-world distribution of \(\{Y^\tau(1),Y^\tau(0)\}\). The parameter of interest is the win odds, defined as 
\[
\theta_h(\tau)
=
\psi_h(\tau)/
\{1-\psi_h(\tau)\}.
\]

\section{Identification and Efficient Influence Function}

We now relate the marginal win estimand defined in Section~2 to the observed data.
We assume consistency: for a subject receiving treatment \(A=a\), the underlying event history under the observed treatment agrees with the corresponding potential event history under treatment \(a\). We also assume conditional exchangeability:
\(
\{Y^\tau(1),Y^\tau(0)\}\perp A\mid X,
\)
and treatment positivity:
\(
0<\pi(X)<1,
\) where \(\pi(X)=P(A=1\mid X).\) In a randomized trial, \(\pi(X)\) is known by design.

Under consistency, we may write the full-data outcome under the observed
treatment arm using the corresponding potential event times. For a scalar
right-censored outcome, let \(T=T(A)\) denote the event time under the observed
treatment arm, and define
\(
Y^\tau=T\wedge\tau.
\)
For a two-component semi-competing risk outcome, let
\(T_1=T_1(A)\) and \(T_2=T_2(A)\) denote the higher-priority and
lower-priority event times under the observed treatment arm, respectively, and
define
\(
T_1^\tau=T_1\wedge\tau,
\
T_2^\tau=T_2\wedge T_1\wedge\tau,
\ \mbox{and}\
Y^\tau=(T_1^\tau,T_2^\tau).
\)

Let \(K\) denote a revelation time, defined as a time by which the restricted data outcome \(Y^\tau\) is fully determined. The revelation time determines the period over which censoring must be accounted for in the observed-data
representation: if a subject is uncensored through \(K\), then the restricted outcome \(Y^\tau\) is observed. For the scalar case, the restricted outcome is fully determined once the event occurs or the time horizon is reached, so
\(
K=T\wedge\tau.
\)
For the two-component semi-competing risk outcome, a valid revelation time is
\(
K=T_1\wedge\tau.
\)
If the terminal event occurs before \(\tau\), then \(T_1^\tau\) is known at \(T_1\), and \(T_2^\tau\) is also determined at \(T_1\) because the lower-priority event is used only up to the terminal event. If the terminal
event does not occur before \(\tau\), then both components are determined at \(\tau\). Thus, censoring must be handled up to \(T_1\wedge\tau\), rather than up to a hypothetical event time of lower priority after the terminal event.

Let \(N^F(t)\) denote the full event-history process in the absence of censoring. In the scalar setting, \(N^F(t)=I(T\le t)\). In the semi-competing risk setting, let
\[
N^F(t)=\{N_1(t),N_2(t)\},
\]
where
\[
N_1(t)=I(T_1\le t),\qquad
N_2(t)=I(T_2\le t,\ T_2<T_1).
\]
Thus, the nonterminal-event process records only events occurring before the terminal event.

Let \(L=(X,A,\{N^F(t):0\le t\le\tau\})\) denote the full data that would be available in the absence of censoring. Define the full-data filtration
\[
\mathcal F_t^F
=\sigma\{X,A,N^F(s):0\le s\le t\},
\]
and let
\[
H^F_{t-}
=\bigl(X,\{N^F(s):0\le s<t\}\bigr)
\]
denote the baseline covariates and the full predictable event
history immediately before time \(t\).  Let \(\{\mathcal F_t^O:t\ge0\}\) denote the filtration of the observed data generated by the baseline covariates, treatment, the event-history processes observed up to time \(t\), and
the censoring process \citep{FlemingHarrington1991}. We write \(H^O_{t-}\) for the corresponding predictable
observed event history. Before censoring and revelation of the restricted outcome, \(H^O_{t-}\) agrees with \(H^F_{t-}\).

In the scalar setting, before the revelation time, the event history contains no additional information beyond survival
up to time \(t\). Consequently, the relevant history for the censoring and residual projections reduces to \(X\),
conditional on remaining at risk. In the semi-competing risk setting, the predictable history includes \(X\) and the
terminal and nonterminal event histories prior to time \(t\). This permits the censoring mechanism and the augmentation
term in the efficient influence function to depend on the evolving event history.

\subsection{Observed Data and Censoring Assumptions}

Let \(C\) denote the censoring time. The observed data vector is denoted by \(O\), with the exact form depending on the outcome process. For a scalar right-censored outcome, one observes
\(
O=(X,A,\widetilde T,\Delta),
\)
where
\(
\widetilde T=T\wedge C\wedge\tau,
\quad
\Delta=I(T\le C,\ T\le\tau).
\)
Thus, the restricted outcome \(Y^\tau=T\wedge\tau\) is fully observed if
and only if the subject is uncensored through \(K=T\wedge\tau\).

For a two-component semi-competing risk outcome, it is more convenient to represent the observed data through event-history processes rather than only through the final restricted variables. This is because censoring may occur
before the revelation time, and both the observed partial event history and the history-dependent censoring mechanism enter the observed-data influence function. Specifically, using the counting processes \(N_1\) and \(N_2\) defined above for the terminal and nonterminal event, the observed data contain
\[
\left\{
N_1(t\wedge C\wedge\tau),
N_2(t\wedge C\wedge\tau):0\le t\le\tau
\right\},
\]
together with \(X\), \(A\), and the observed censoring information. In this setting, the restricted outcome \(Y^\tau=(T_1^\tau,T_2^\tau)\) is fully observed if and only if the subject is uncensored through the revelation
time \(K=T_1\wedge\tau\). In both outcome settings, define
\(
R=I(C\ge K).
\)
Thus, \(R=1\) indicates that \(Y^\tau\) is fully observed. In the scalar case, \(R\) is distinct from the event indicator \(\Delta\): subjects who remain event-free and uncensored through \(\tau\) have \(R=1\) and
\(\Delta=0\).

We assume sequential independent censoring, formulated as a coarsening-at-random condition relative to the full event history \(L\), not merely as a restriction on the censoring intensity in the observed-data
filtration \citep{vanderLaanRobins2003,Tsiatis2006}. For simplicity, assume that censoring has an absolutely continuous conditional hazard on the period before revelation and that censoring and event jumps do not occur
simultaneously. For each \(a\in\{0,1\}\), suppose that, for almost every \(0<u<K\),
\begin{equation}
\lim_{\delta\downarrow0}\frac{1}{\delta}
P\bigl(u\le C<u+\delta\mid C\ge u,L,A=a\bigr)
=\lambda_a^C(u\mid H^F_{u-}).
\label{eq:car_full_event_history}
\end{equation}
The right-hand side depends only on the predictable event history before \(u\), rather than on subsequent event outcomes. On the observed risk set \(\{C\ge u,K\ge u\}\), this history is available from the observed data.

Define the cumulative censoring hazard and corresponding survival weight along the event-history path by
\begin{equation}
\Lambda_a^C(t;L)
=\int_0^t\lambda_a^C(u\mid H^F_{u-})\,du,
\qquad
G_a(t;L)=\exp\{-\Lambda_a^C(t;L)\},
\quad 0\le t\le K.
\label{eq:censoring_path_weight}
\end{equation}
Under \eqref{eq:car_full_event_history},
\(
G_a(t;L)=P(C\ge t\mid L,A=a)
\)
for \(0\le t\le K\). The notation \(G_a(t;L)\) emphasizes that the censoring weight is constructed from the event-history path through \(t\). In general, for an internal time-dependent history, it should
not be identified with \(P(C\ge t\mid A=a,H^F_{t-})\) without additional justification. We assume censoring positivity: there exists a constant \(\varepsilon>0\) such that
\[
G_a(t;L)\ge\varepsilon,
\qquad 0\le t\le K,\qquad a\in\{0,1\},
\]
almost surely.

These assumptions yield the inverse-probability-of-censoring identity
\begin{equation}
E\left\{
\left.\frac{R\,f(Y^\tau,X)}{G_a(K;L)}\right|A=a,X
\right\}
=E\{f(Y^\tau,X)\mid A=a,X\}
\label{eq:conditional_ipcw_identification}
\end{equation}
for any integrable measurable function \(f\), where, conditional on the full event history, \(E(R\mid L,A=a)=G_a(K;L)\). The weighted quantity in \eqref{eq:conditional_ipcw_identification} is observable, since
\(G_a(K;L)\) is evaluated only for subjects with \(R=1\), whose event histories are observed through \(K\). Together with consistency, conditional exchangeability, and treatment positivity, this identity
identifies the marginal distributions \(F_a^\tau\) of \(Y^\tau(a)\), and hence
\[
\psi_h(\tau)
=\int\!\!\int h(y_1,y_0)\,
dF_1^\tau(y_1)\,dF_0^\tau(y_0).
\]
This formulation does not require the joint cross-world distribution of \(\{Y^\tau(1),Y^\tau(0)\}\).

\subsection{Nuisance Functions}

The efficient influence function depends on several nuisance functions.
First, define the kernel projection functions
\[
h_1(y)
=
E\left[h\{y,Y^\tau(0)\}\right],
\quad \mbox{and}\quad
h_0(y)
=
E\left[h\{Y^\tau(1),y\}\right].
\]
These are the first-order projections of the pairwise comparison kernel
with respect to the distributions of the marginal potential outcome of the control and treatment, respectively. Next, define the covariate projection functions
\[
\mu_1(X)
=
E\left[h_1\{Y^\tau(1)\}\mid X\right],
\quad \mbox{and}\quad
\mu_0(X)
=
E\left[h_0\{Y^\tau(0)\}\mid X\right].
\]
For \(a\in\{0,1\}\), define the centered residuals
\[
\zeta_1(Y^\tau,X)=h_1(Y^\tau)-\mu_1(X),
\quad \mbox{and}\quad
\zeta_0(Y^\tau,X)=h_0(Y^\tau)-\mu_0(X).
\]
By consistency and conditional exchangeability,
\(E\{\zeta_a(Y^\tau,X)\mid X,A=a\}=0\). Define the observed censoring counting process and at-risk process by
\[
N^C(t)=I(C\le t,\ C<K),
\quad \mbox{and}\quad
Y^C(t)=I(C\ge t,\ K\ge t).
\]
For treatment arm \(a\), let \(\lambda_a^C(u\mid H^F_{u-})\) be the conditional censoring hazard specified in Section~3.1. Its cumulative hazard along the full event-history path is
\[
\Lambda_a^C(t;L)
=
\int_0^t\lambda_a^C(u\mid H^F_{u-})\,du,
\qquad 0\le t\le K,
\]
and the corresponding censoring survival weight is
\(G_a(t;L)=\exp\{-\Lambda_a^C(t;L)\}\).
For a subject with \(A=a\), define the censoring martingale
\[
M_a^C(t)
=
N^C(t)
-
\int_0^t Y^C(u)\,
\lambda_a^C(u\mid H^O_{u-})\,du,
\qquad 0\le t\le\tau.
\]
On the at-risk set \(\{Y^C(u)=1\}\), the full and observed predictable event histories coincide, so the compensator can
equivalently be written using \(H^F_{u-}\). Under the sequential independent censoring assumption,
\(M_a^C\) is a martingale with respect to the observed-data filtration for subjects in arm \(a\)
\citep{FlemingHarrington1991}.

Finally, define the predictable residual projection
\[
q_a(u,H^F_{u-})
=
E\left\{
\zeta_a(Y^\tau,X)
\mid K\ge u,\ H^F_{u-},\ A=a
\right\},
\qquad 0\le u\le\tau,
\]
on the event \(\{K\ge u\}\). This is the conditional mean of the full-data residual among subjects whose restricted outcome has not yet been revealed, given their predictable event history.
The projection uses the full relevant history for the nonparametric efficiency calculation. It is evaluable from the observed history while a subject remains uncensored and at risk. In the scalar
right-censored setting, before revelation this reduces to \(q_a(u,X)=E\{\zeta_a(Y^\tau,X)\mid K\ge u,X,A=a\}\).

\subsection{Observed-Data Efficient Influence Function}

For simplicity, write $\psi=\psi_h(\tau)$, and let
\(
\eta=(\pi,G,h_1,h_0,\mu_1,\mu_0,q)
\)
denote the collection of population nuisance parameters. The censoring hazard entering the martingale $M_a^C$ is equivalently specified through $G_a$. We work under the consistency, conditional exchangeability,
sequential independent censoring (coarsening-at-random), and positivity
assumptions of Section~3.1. In particular, for some $\epsilon>0$,
\[
\epsilon\le\pi(X)\le1-\epsilon,
\qquad
G_a(u;L)\ge\epsilon,
\quad 0\le u\le K,\quad a\in\{0,1\},
\]
almost surely. As specified in Section~3.1, censoring hazards are absolutely continuous and censoring and event processes have no simultaneous jumps. We assume that all relevant conditional expectations
are well defined and that the influence-function expression below is
square integrable. For the efficiency assertion, we consider a nonparametric event-history model with otherwise unrestricted predictable event-transition intensities on their active risk sets, and assume the
predictable martingale representation property for the natural observed-data
filtration. The residual projection $q_a$ conditions on the full relevant predictable event history.

The following theorem characterizes the observed-data efficient influence function for the marginal win probability. Its proof is given in the Supplementary Material.

\begin{thm}
\label{thm:eif}
For $a\in\{0,1\}$, define
\[
\phi_a(O;\eta)
=
\frac{R}{G_a(K;L)}\zeta_a(Y^\tau,X)
+
\int_0^{K\wedge C}
\frac{q_a(u,H^O_{u-})}{G_a(u;L)}\,dM_a^C(u),
\]
for \(a=0,\ 1\), where the censoring weights and residual projections are evaluated using
the observed event history while the subject is at risk. Then the  efficient influence function for $\psi$ is
\begin{equation}
\label{eq:eif_observed_corrected}
\phi_{\mathrm{obs}}(O;\psi,\eta)
=\mu_1(X)+\mu_0(X)-2\psi
+\frac{A}{\pi(X)}\phi_1(O;\eta)
+\frac{1-A}{1-\pi(X)}\phi_0(O;\eta).
\end{equation}
Consequently, the semiparametric efficiency bound for the asymptotic
variance of a regular estimator of $\psi$ is
\[
\mathcal I_{\mathrm{eff}}^{-1}(\psi)
=E\bigl\{\phi_{\mathrm{obs}}^2(O;\psi,\eta)\bigr\}.
\]
\end{thm}

The efficient influence function can be decomposed as
\[
\phi_{\mathrm{obs}}(O;\psi,\eta)
=\phi_{\mathrm{cov}}(O)+\phi_{\mathrm{IPCW}}(O)+\phi_{\mathrm{aug}}(O),
\]
where
\(
\phi_{\mathrm{cov}}(O)=\mu_1(X)+\mu_0(X)-2\psi
\)
is the covariate projection term inherited from the full-data causal
U-statistic \citep{Mao2018},
\[
\phi_{\mathrm{IPCW}}(O)
=\frac{A}{\pi(X)}\frac{R}{G_1(K;L)}\zeta_1(Y^\tau,X)
+\frac{1-A}{1-\pi(X)}\frac{R}{G_0(K;L)}\zeta_0(Y^\tau,X)
\]
is the inverse-probability-of-censoring weighted complete-outcome
contribution, and
\[
\phi_{\mathrm{aug}}(O)
=\frac{A}{\pi(X)}\int_0^{K\wedge C}
\frac{q_1(u,H^O_{u-})}{G_1(u;L)}\,dM_1^C(u)
+\frac{1-A}{1-\pi(X)}\int_0^{K\wedge C}
\frac{q_0(u,H^O_{u-})}{G_0(u;L)}\,dM_0^C(u)
\]
is the censoring-martingale augmentation term.

The IPCW contribution uses subjects whose restricted outcomes are fully
revealed, while the augmentation uses observed event-history information
prior to censoring. The predictable residual projection $q_a$ is chosen so
that the augmentation cancels the covariance of the IPCW term with
censoring nuisance scores.  In the scalar right-censored setting, the relevant history
contains no information beyond baseline covariates and survival up to the current time, so the projection $q_a(u,X)$ depends only on time and baseline covariates.  In the semi-competing risk setting, it also includes the
evolving terminal and nonterminal event histories. Hence event-history
information may improve precision while the target estimand remains the
marginal win probability $\psi_h(\tau)$.

Finally, for the win odds $\theta=\psi/(1-\psi)$, where $0<\psi<1$, the delta method gives the efficient influence function
\[
\phi_\theta(O)
=\frac{\phi_{\mathrm{obs}}(O;\psi,\eta)}{(1-\psi)^2}.
\]

\section{Estimation and Asymptotic Properties}

\subsection{Cross-Fitted Nuisance Estimation}

We construct a cross-fitted one-step estimator by combining
the fold-specific plug-in win probability from \eqref{eq:cf_plugin_psi} with an empirical correction based
on the efficient influence function in Theorem~\ref{thm:eif}. The one-step estimator follows the general principle of efficient estimation based on influence functions \citep{vanderVaart1998}, while cross-fitting
\citep{Chernozhukov2018} allows flexible nuisance estimation without imposing Donsker conditions. Let
$\mathcal I_1,\ldots,\mathcal I_V$ partition $\{1,\ldots,n\}$, and write
$\mathcal T_v=\{1,\ldots,n\}\setminus\mathcal I_v$ for the training
indices of fold $v$. All nuisance functions carrying the superscript $(-v)$ are fitted using $\mathcal T_v$ and evaluated on $\mathcal I_v$. In a randomized trial, the treatment probability $\pi(X)$ is known
by design and need not be estimated.

Within treatment arm $a$, let $\widehat\lambda_a^{C,(-v)}(u\mid H^O_{u-})$
denote an estimated censoring hazard. Its corresponding survival weight
along the observed event-history path is
\[
\widehat G_a^{(-v)}(t;L)
=\exp\left\{-\int_0^t
\widehat\lambda_a^{C,(-v)}(u\mid H^O_{u-})\,du\right\},
\qquad t\le K.
\]
Although the notation uses $L$, the estimated survival weight is evaluated only along the observed event-history
path before censoring or revelation. Thus, $\widehat G_a^{(-v)}(t;L)$ is computable for
$t\le K\wedge C$. In particular, evaluation at $K$ is required only for subjects with $R=1$,
for whom $K$ and the event history through $K$ are observed.

For a scalar right-censored outcome, the censoring model conditions on baseline covariates and time. An arm-specific Kaplan--Meier estimator is appropriate when censoring does not depend on $X$ within that arm,
whereas a Cox model or other conditional survival method can accommodate covariate dependence. For semi-competing risks, the censoring hazard may depend on predictable event-history variables, such as prior nonterminal event status and time since that event.

In our numerical implementation, the event and censoring hazards are estimated using penalized maximum likelihood within each training fold. We consider parametric working hazard models as well as more flexible specifications that allow flexible baseline hazard functions through penalized cubic B-splines. Ridge regularization is used to stabilize the regression coefficients, while the spline coefficients are additionally regularized through a second-difference penalty. These regularization methods help prevent extreme fitted hazard rates and excessively small estimated censoring survival probabilities, which can lead to unstable inverse-censoring weights and contributions to influence functions. The specific working models and tuning parameters are described in the numerical studies. These modeling and regularization choices are not requirements of the proposed estimation procedure, and the asymptotic results apply when the nuisance estimators satisfy the stated  conditions on consistency and rate of convergence.

To estimate the marginal outcome distributions, put $\pi_1(X)=\pi(X)$ and $\pi_0(X)=1-\pi(X)$.
Use the known randomization probabilities where applicable, or training-fold estimates $\widehat\pi_a^{(-v)}$ otherwise. Define
\[
\widehat W_{a,j}^{(-v)}
=\frac{I(A_j=a)R_j}
{\widehat\pi_a^{(-v)}(X_j)\widehat G_a^{(-v)}(K_j;L_j)},
\qquad j\in\mathcal T_v,
\]
with the convention that the contribution is zero when $R_j=0$, so neither an unobserved $Y_j^\tau$ nor an unobserved $K_j$ is needed. The normalized IPCW empirical distribution is
\begin{equation}
\label{eq:cf_marginal_F}
\widehat F_a^{\tau,(-v)}(B)
=\frac{\sum_{j\in\mathcal T_v}\widehat W_{a,j}^{(-v)}I(Y_j^\tau\in B)}
       {\sum_{j\in\mathcal T_v}\widehat W_{a,j}^{(-v)}},
\qquad a\in\{0,1\},
\end{equation}
provided that the denominator is positive. The inverse treatment probability
factor cancels in this normalized expression if randomization is constant within arm.

We construct the kernel projections from these same fitted marginal distributions:
\[
\widehat h_1^{(-v)}(y)
=\int h(y,y_0)\,d\widehat F_0^{\tau,(-v)}(y_0),\ \mbox{and}\
\widehat h_0^{(-v)}(y)
=\int h(y_1,y)\,d\widehat F_1^{\tau,(-v)}(y_1).
\]
The corresponding fold-specific plug-in win probability is
\begin{equation}
\label{eq:cf_plugin_psi}
\widehat\psi_{\mathrm{pl}}^{(-v)}
=\iint h(y_1,y_0)\,
 d\widehat F_1^{\tau,(-v)}(y_1)
 d\widehat F_0^{\tau,(-v)}(y_0).
\end{equation}
This coherent construction of $\widehat\psi_{\mathrm{pl}}$ and $\widehat h_a$ is important for cancellation of the first-order kernel-projection errors in the subsequent one-step analysis.

Within treatment arm $a$, the covariate projection $\mu_a(X)$ may be estimated by an IPCW-adjusted regression of the pseudo-outcome $\widehat h_a^{(-v)}(Y^\tau)$ on $X$. The predictable residual
projection is estimated using the corresponding residual $\widehat h_a^{(-v)}(Y^\tau)-\widehat\mu_a^{(-v)}(X)$.
A risk-set regression for $q_a(u,H^O_{u-})$ must account for the fact
that the future restricted outcome may itself be censored, and fitting such
a regression using only fully observed outcomes without an appropriate censoring adjustment is generally biased. One option is an IPCW-adjusted risk-set regression. For example, among fully observed outcomes, weighting
by the inverse probability of remaining uncensored through the revelation time $K$ accounts for
censoring of the future restricted outcome. The regression must also condition on the relevant
at-risk event history at time $u$. Another approach is to use fitted models for the event-history
process to evaluate the conditional expectation of the residual. The remaining event trajectory is generated or integrated over conditional on $K\ge u$ and the available history, and the resulting residuals are averaged. The conditional expectation may be approximated using Monte Carlo simulation or deterministic numerical integration.

The theoretical efficient influence function conditions on the full relevant predictable event history. A lower-dimensional history summary may be used computationally when it is sufficiently informative
for the required conditional projection, and otherwise an efficiency claim
based on the nonparametric model of Theorem~\ref{thm:eif} is not justified without further assumptions. Similarly, setting $q_a=0$ yields a simpler working estimator but does not generally attain the
semiparametric efficiency bound. When the censoring hazard is estimated, it can also lose the nuisance-error cancellation required by the one-step asymptotic theory. The convergence conditions for the chosen
nuisance estimators are stated in the asymptotic theory below.

\subsection{Cross-Fitted One-Step Estimator}
\label{subsec:one_step}

We combine the fold-specific plug-in win probability from \eqref{eq:cf_plugin_psi} with an empirical correction based on
the efficient influence function in Theorem~\ref{thm:eif}. For each fold $v$, let
$\widehat\psi_v=\widehat\psi_{\mathrm{pl}}^{(-v)}$
denote the plug-in estimator constructed from the estimated
marginal outcome distributions using the training sample
$\mathcal T_v$. For a validation observation $i\in\mathcal I_v$, all fitted quantities below are obtained
from the same training sample. Define, for $a\in\{0,1\}$,
\[
\widehat\zeta_{a,i}^{(-v)}
=\widehat h_a^{(-v)}(Y_i^\tau)
 -\widehat\mu_a^{(-v)}(X_i),
\]
where this residual is evaluated only when $R_i=1$ and $A_i=a$. The kernel projections $\widehat h_a^{(-v)}$ are constructed from the same estimated marginal outcome distributions as $\widehat\psi_v$, as specified in Section~4.1.

For $A_i=a$, define the estimated censoring-martingale increment by
\[
d\widehat M_{a,i}^{C,(-v)}(u)
=dN_i^C(u)-Y_i^C(u)
\widehat\lambda_a^{C,(-v)}(u\mid H^O_{u-,i})\,du,
\]
and write $\widehat G_{a,i}^{(-v)}(u) =\widehat G_a^{(-v)}(u;L_i)$ for the estimated censoring
survival weight along the observed history of subject $i$. The arm-specific contribution is
\begin{equation}
\label{eq:estimated_arm_contribution}
\widehat\phi_{a,i}^{(-v)}
=\frac{R_i}{\widehat G_{a,i}^{(-v)}(K_i)}
\widehat\zeta_{a,i}^{(-v)}
+\int_0^{K_i\wedge C_i}
\frac{\widehat q_a^{(-v)}(u,H^O_{u-,i})}
{\widehat G_{a,i}^{(-v)}(u)}
\,d\widehat M_{a,i}^{C,(-v)}(u),
\qquad A_i=a.
\end{equation}
The martingale integral is computed from the observed event and censoring histories up to $K_i\wedge C_i$. Although $\widehat M_{a,i}^{C,(-v)}$ is written in martingale form,
it need not be a martingale under the true law when the censoring hazard is estimated.

For each $i\in\mathcal I_v$, the estimated influence-function contribution is
\begin{equation}
\label{eq:estimated_eif_one_step}
\widehat\phi_{\mathrm{obs},i}^{(-v)}
=\widehat\mu_1^{(-v)}(X_i)
 +\widehat\mu_0^{(-v)}(X_i)-2\widehat\psi_v
+\frac{A_i}{\widehat\pi^{(-v)}(X_i)}
  \widehat\phi_{1,i}^{(-v)}
 +\frac{1-A_i}{1-\widehat\pi^{(-v)}(X_i)}
  \widehat\phi_{0,i}^{(-v)},
\end{equation}
where only the arm-specific term corresponding to $A_i$ is evaluated. The proposed cross-fitted one-step estimator is
\begin{equation}
\label{eq:cross_fitted_one_step}
\widehat\psi_{\mathrm{os}}
=\sum_{v=1}^V\frac{n_v}{n}
\left\{\widehat\psi_v
+\frac{1}{n_v}\sum_{i\in\mathcal I_v}
\widehat\phi_{\mathrm{obs},i}^{(-v)}\right\}.
\end{equation}
A direct alternative is to substitute the fitted nuisance functions into the influence-function
estimating equation and solve for $\psi$. However, because the kernel projections $h_1$
and $h_0$ depend on the unknown marginal outcome distributions, treating their estimates as fixed
in this equation generally leaves first-order errors from estimation of these distributions.
The proposed one-step estimator instead combines a coherent plug-in estimate of $\psi$ with an
empirical influence-function correction. As shown in the Supplementary Material,
the first-order kernel-projection errors cancel, leaving a second-order remainder involving the
product of the two marginal-distribution estimation errors. This cancellation is important
for establishing asymptotic efficiency under the nuisance-estimation conditions in Section~4.3.

For comparison, we use a normalized inverse-probability-of-censoring weighted (IPCW) two-sample statistic, in the spirit of existing methods for censored win statistics
\citep{Dong2020IPCW,Dong2021Dependent}. Define the cross-fitted individual weights
\[
\widehat w_{a,i}
=\frac{I(A_i=a)R_i}
{\widehat\pi_a^{(-v(i))}(X_i)
 \widehat G_a^{(-v(i))}(K_i;L_i)},
\quad
\widehat\pi_1=\widehat\pi,\quad
\widehat\pi_0=1-\widehat\pi,
\]
where a weight is zero when $R_i=0$. The comparator is
\begin{equation}
\label{eq:ipcw_comparator}
\widehat\psi_{\mathrm{IPCW}}
=\displaystyle\sum_{i=1}^n\sum_{j=1}^n
\widehat w_{1,i}\widehat w_{0,j}
 h(Y_i^\tau,Y_j^\tau)\bigg/\displaystyle\left(\sum_{i=1}^n\widehat w_{1,i}\right)
 \left(\sum_{j=1}^n\widehat w_{0,j}\right).
\end{equation}
The comparison uses only pairs whose restricted outcomes are both fully revealed. For constant treatment probabilities within arms, the inverse treatment-probability factors cancel after normalization.
The IPCW estimator provides a reference for assessing the precision gains from baseline covariate adjustment and the censoring-martingale augmentation.

\subsection{Asymptotic Normality and Efficiency}
\label{subsec:one_step_asymptotics}

Write $Pf=E\{f(O)\}$, $\mathbb P_{n,v}f=n_v^{-1}\sum_{i\in\mathcal I_v}f(O_i)$,
and $p_a(X)=P(A=a\mid X)$, for $a=0,1$.  For each validation fold $v$,
all the fitted nuisance functions and the plug-in probability $\widehat\psi_v=\widehat\psi_{\mathrm{pl}}^{(-v)}$ are trained only on $\mathcal T_v$.  Write $F_a=F_a^\tau$, $\widehat F_{a,v}=\widehat F_a^{\tau,(-v)}$, and
$\Delta_{a,v}=\widehat F_{a,v}-F_a$.  The fitted kernel projections $\widehat h_{1,v}$ and $\widehat h_{0,v}$ and the plug-in probability $\widehat\psi_v$ are constructed from the same two fitted marginal distributions, as in Section~4.1.

We assume the following regularity conditions, in addition to the identification and model conditions of Theorem~\ref{thm:eif}.

\begin{enumerate}
\item[(C1)] The observations are i.i.d.; the number $V$ of validation folds is fixed, and $n_v/n\to\rho_v\in(0,1)$ for every $v$. The kernel satisfies $0\le h\le1$, the horizon $\tau$ is finite,
and $0<\sigma^2=E\{\phi_{\mathrm{obs}}^2(O;\psi,\eta)\}<\infty$.
The true and fitted treatment probabilities and censoring survival weights are uniformly bounded away from zero.  The true and fitted predictable censoring hazards are uniformly bounded.  The fitted
survival weights are obtained by integrating the corresponding fitted
hazards.  The fitted regressions are bounded (or clipped) to their natural ranges. The validation-fold partition is fixed or generated independently of the observations.

\item[(C2)] For each fold, conditional on its training sample, let
\[
\begin{aligned}
 e_{p,a,v}&=\|\widehat p_{a,v}-p_a\|_{2,P_X},&
 e_{h,a,v}&=\|\widehat h_{a,v}-h_a\|_{2,F_a},\\
 e_{\mu,a,v}&=\|\widehat\mu_{a,v}-\mu_a\|_{2,P_X},&
 e_{q,a,v}&=\|\widehat q_{a,v}-q_a\|_{2,\nu_a},\\
 e_{\lambda,a,v}&=\|\widehat\lambda_{a,v}^C-\lambda_a^C\|_{2,\nu_a},
\end{aligned}
\]
where $\|f\|_{2,\nu_a}^2=E_a\{\int_0^K f^2(u,H^F_{u-})\,du\}$, and 
$E_a$ integrates the baseline distribution and arm $a$ uncensored
event trajectory.  For every $a,v$, all five errors are $o_p(1)$,
and $|\widehat\psi_v-\psi|=o_p(1)$.

\item[(C3)] The bilinear kernel remainder satisfies
\begin{equation}
\label{eq:thm2_bilinear_rate}
\max_{1\le v\le V}\left|
\iint h(y_1,y_0)\,d\Delta_{1,v}(y_1)d\Delta_{0,v}(y_0)
\right|=o_p(n^{-1/2}).
\end{equation}

\item[(C4)] Uniformly over the finitely many folds,
\begin{equation}
\label{eq:thm2_product_rates}
\sum_{a=0}^1 e_{p,a,v}\{e_{h,a,v}+e_{\mu,a,v}\}
+\sum_{a=0}^1 e_{\lambda,a,v}
\{e_{q,a,v}+e_{h,a,v}+e_{\mu,a,v}\}
=o_p(n^{-1/2}).
\end{equation}
When $p_a$ is known by randomization and used in estimation,
$e_{p,a,v}=0$ and the first sum vanishes.
\end{enumerate}

\begin{thm}
\label{thm:asymptotic_normality}
Under the conditions of Theorem~\ref{thm:eif} and \textup{(C1)--(C4)}, the cross-fitted one-step estimator in \eqref{eq:cross_fitted_one_step} is asymptotically linear:
\[
\sqrt n(\widehat\psi_{\mathrm{os}}-\psi)
=\frac1{\sqrt n}\sum_{i=1}^n
\phi_{\mathrm{obs}}(O_i;\psi,\eta)+o_p(1).
\]
Consequently,
\[
\sqrt n(\widehat\psi_{\mathrm{os}}-\psi)
\overset{d}{\longrightarrow}N(0,\sigma^2),\qquad
\sigma^2=E\{\phi_{\mathrm{obs}}^2(O;\psi,\eta)\}.
\]
Thus the estimator attains the semiparametric efficiency bound in Theorem~\ref{thm:eif}.
\end{thm}

The proof is given in the Supplementary Material. The joint construction of the plug-in probability
and kernel projections ensures cancellation of their first-order estimation errors. With known, uniformly bounded treatment and censoring weights, condition~\textup{(C3)} holds for the normalized
IPCW empirical distributions. With estimated weights, additional rate or stability conditions
may be needed, and outer cross-fitting alone does not guarantee~\textup{(C3)}.

\subsection{Inference for Win Odds}
\label{subsec:inference_win_odds}

Under the conditions of Theorem~\ref{thm:asymptotic_normality}, the asymptotic variance of $\sqrt (\widehat\psi_{\mathrm{os}}-\psi)$ is $\sigma^2=E\{\phi_{\mathrm{obs}}^2(O;\psi,\eta)\}$. Let
$\widehat\phi_{\mathrm{obs},i}^{(-v)}$ denote the cross-fitted influence-function contribution defined in
\eqref{eq:estimated_eif_one_step} for $i\in\mathcal I_v$, and write 
\[
\overline{\widehat\phi}_{\mathrm{obs}}
=\frac1n\sum_{v=1}^V\sum_{i\in\mathcal I_v}
\widehat\phi_{\mathrm{obs},i}^{(-v)}.
\]
We estimate $\sigma^2$ by the sample variance of these contributions,
\begin{equation}
\label{eq:one_step_variance}
\widehat\sigma^2
=\frac1n\sum_{v=1}^V\sum_{i\in\mathcal I_v}
\left\{\widehat\phi_{\mathrm{obs},i}^{(-v)}
-\overline{\widehat\phi}_{\mathrm{obs}}\right\}^2.
\end{equation}
The estimated nuisance functions, whether obtained from parametric or spline-based hazard models, are incorporated into the estimated influence-function contributions and do not alter the form of the
variance estimator in \eqref{eq:one_step_variance}. Under conditions~\textup{(C1)--(C2)},
$\widehat\sigma^2\overset{p}{\longrightarrow}\sigma^2$. This follows from the $L_2(P)$ consistency of the cross-fitted
influence-function contributions, as established in the Supplementary Material. The centering in
\eqref{eq:one_step_variance} is retained because the one-step estimator does not generally solve an estimating equation whose empirical mean is zero.

An approximate $100(1-\alpha)\%$ Wald confidence interval for the marginal win probability is
\[
\widehat\psi_{\mathrm{os}}
\ \pm\ z_{1-\alpha/2}\frac{\widehat\sigma}{\sqrt n}.
\]

When $0<\widehat\psi_{\mathrm{os}}<1$, the estimator of the win odds \(\theta=\psi/(1-\psi)\) is
\[
\widehat\theta_{\mathrm{os}}
=\frac{\widehat\psi_{\mathrm{os}}}
{1-\widehat\psi_{\mathrm{os}}}.
\]
By the delta method,
\[
\sqrt n(\widehat\theta_{\mathrm{os}}-\theta)
\overset{d}{\longrightarrow}
N\left(0,\frac{\sigma^2}{(1-\psi)^4}\right).
\]
The standard deviation is estimated by
\[
\widehat{\mathrm{SE}}(\widehat\theta_{\mathrm{os}})
=\frac{\widehat\sigma}
{\sqrt n\,(1-\widehat\psi_{\mathrm{os}})^2}.
\]
For inference on the log win odds, write $\ell=\log\theta=\log\{\psi/(1-\psi)\}$. Again by the delta method,
\[
\sqrt n\{\log(\widehat\theta_{\mathrm{os}})-\ell\}
\overset{d}{\longrightarrow}
N\left(0,\frac{\sigma^2}{\psi^2(1-\psi)^2}\right).
\]
An approximate $100(1-\alpha)\%$ confidence interval for the win odds is therefore obtained by exponentiating
\begin{equation}
\label{eq:win_odds_log_interval}
\log(\widehat\theta_{\mathrm{os}})
\ \pm\ z_{1-\alpha/2}
\frac{\widehat\sigma}
{\sqrt n\,\widehat\psi_{\mathrm{os}}
(1-\widehat\psi_{\mathrm{os}})}.
\end{equation}
The log-scale construction yields a positive confidence interval for $\theta$. The transformations above are applied when the estimated win probability lies in $(0,1)$, an event with probability tending to one when $0<\psi<1$ and the one-step estimator is consistent.

For the IPCW estimator, Wald inference is based on a separate first-order influence-function approximation that accounts for estimation of the censoring distribution. The corresponding influence-function contributions include the weighted two-sample components and a correction based on the censoring-model score and information matrix. In our numerical implementation, the censoring hazard is estimated using penalized likelihood, with a pooled model that includes treatment and baseline covariates and permits time-varying baseline hazards. The empirical variance of the resulting influence-function contributions is used to construct confidence intervals. Unlike the proposed one-step estimator, this IPCW variance calculation relies on a first-order approximation for the fitted censoring model and may be sensitive to model misspecification.

\section{Simulation Study}
\subsection{Simulation Settings}

We conducted simulation studies to evaluate the finite-sample performance of the proposed cross-fitted one-step estimator of the marginal win probability. We assessed empirical bias,
variability, accuracy of variance estimation, coverage of confidence intervals, and efficiency relative to the IPCW estimator described in Section 4.  We considered both a scalar right-censored time-to-event outcome and a two-component prioritized semi-competing risk outcome. In each setting, the total sample size was $n=200$, $400$, or
$800$, with treatment randomized according to $A\sim\operatorname{Bernoulli}(0.5)$. The baseline covariate $X$ was generated from a standard normal distribution truncated to $[-2,2]$. Each configuration was evaluated using 5000 Monte Carlo replications, with analysis horizon $\tau=5$.

For the scalar setting, the event time was supposed to follow an exponential distribution with hazard
$\lambda_T(A,X)=\lambda_0\exp(\beta_A A+\beta_X X)$, where $\lambda_0=0.10$ and $\beta_A=-0.40$.
We considered three scenarios, labeled ``moderate'', ``stronger'', and ``strongest'', with increasingly strong prognostic effects of the baseline covariate $X$ and varying degrees of censoring.
The corresponding values of $\beta_X$ were $0.75$, $1.00$, and $1.75$, respectively.
Censoring was generated independently of the event time conditional on $X$, with hazard
$\lambda_C(X)=c_0\exp(\gamma_X X)$. The values of $\gamma_X$ were $0.65$, $0.80$, and $0.80$,
and $c_0$ was calibrated to yield approximately 30\%, 40\%, and 40\% censoring before the revelation
time $K=T\wedge\tau$, respectively.

For the semi-competing risk setting, let $T_1$ denote the terminal event time and $T_2$ the nonterminal event time. The nonterminal event hazard was $\lambda_2(A,X)=\lambda_{20}\exp(\beta_{2A}A+\beta_{2X}X)$,
where $\lambda_{20}=0.20$, $\beta_{2A}=-0.30$, and $\beta_{2X}=1.25$. The terminal event hazard depended on
the nonterminal event history through
$\lambda_1(t\mid A,X,N_2(t-))=\lambda_{10}\exp\{\beta_{1A}A+\beta_{1X}X+\beta_{1N2}N_2(t-)\}$,
where $N_2(t-)=I(T_2<t)$, $\lambda_{10}=0.10$, $\beta_{1A}=-0.35$, and $\beta_{1X}=1.25$.
The three scenarios used $\beta_{1N2}=1.25$, $1.60$, and $2.00$, respectively, allowing the effect of prior nonterminal events on the terminal event hazard to vary.

Censoring in the semi-competing risk setting depended on the evolving event history, with hazard $\lambda_C(t\mid X,N_2(t-)) =c_0\exp\{\gamma_X X+\gamma_{N2}N_2(t-)\}$. The pairs $(\gamma_X,\gamma_{N2})$ were $(0.80,1.25)$, $(0.50,0.50)$, and $(0.40,0.40)$ in the moderate, stronger, and strongest scenarios, respectively. The constant $c_0$ was calibrated separately to yield approximately 30\%, 40\%, and 50\% censoring before $K$. In deciding win or loss, larger event times were considered more favorable.

The proposed estimator was implemented using five-fold cross-fitting. In the scalar setting, event hazards were estimated separately by treatment arm, while the censoring hazard was estimated using a pooled exponential model
including treatment and $X$. In the semi-competing risk setting, the terminal event, nonterminal event, and
censoring hazards were estimated separately within each treatment arm. The terminal event and censoring models
included both $X$ and the time-dependent nonterminal event indicator, whereas the nonterminal event model
included $X$. All hazard models were estimated using ridge-penalized maximum likelihood, with penalty
parameter $\kappa=2$ applied to the regression coefficients but not the intercept.

Within each training fold, the marginal outcome distributions and kernel projections $h_1$ and $h_0$ were estimated using normalized IPCW empirical distributions. The conditional means $\mu_a(X)$ and predictable residual projections $q_a(u,H_{u-})$ were evaluated using the fitted event models. In the scalar setting, these conditional expectations and the censoring augmentation were evaluated analytically. In the semi-competing risk setting, the conditional
expectations were approximated using deterministic conditional quantile integration with 256 grid points,
and the censoring augmentation was evaluated using eight-point Gauss--Legendre quadrature, separately before and after the nonterminal event.

We compared the proposed one-step estimator with the normalized IPCW estimator. We also considered an oracle hazard version of the proposed estimator, which used the true event and censoring hazards while retaining estimation of the marginal outcome distributions and kernel projections. This estimator using oracle hazard served as a benchmark for assessing the effect of nuisance hazard estimation on finite-sample performance.

The true value of the win probability was calculated by numerical integration using 80-point Gaussian quadrature
in the scalar setting. In the semi-competing risk setting, it was approximated using 500,000 independent pairs of
uncensored outcomes generated from the treatment and control distributions. For each estimator, we report
empirical bias, empirical standard deviation,  average estimated standard error and empirical coverage of nominal 95\% confidence intervals for the win probability.  For IPCW, we used a first-order influence-function approximation accounting for estimation of the censoring model. Relative efficiency was defined as the empirical variance of IPCW divided by that of the proposed full-$q_a$ estimator. All simulations were implemented in R.

\subsection{Results}

Tables~\ref{tab:sim_scalar} and \ref{tab:sim_practical} summarize results from 5000 Monte Carlo replications for each scenario. Across both outcome settings, the proposed estimator had small empirical
bias, and its  standard errors based on the influence function were close to the empirical standard deviations.
The empirical coverage rates of 95\% confidence intervals ranged from 94.6\% to 95.6\% in the scalar setting and from 94.5\% to 95.3\% in the semi-competing risk setting. 

For scalar right-censored outcomes, the proposed estimator was more efficient than IPCW across all three scenarios and sample sizes. Relative efficiency, defined as the ratio of the empirical variance of IPCW to that of the proposed estimator, ranged from 1.08 to 2.04. In the ``moderate'' scenario, it was 1.15--1.19, while in the ``strongest''
scenario it ranged from 1.85 to 2.04. The efficiency gains were generally smaller at $n=200$, particularly in the ``stronger'' scenario, where relative efficiency was 1.08, compared with 1.33 and 1.35
at $n=400$ and $800$, respectively. At $n=200$, the oracle hazard benchmark achieved a relative efficiency of 1.47, suggesting that nuisance hazard estimation contributed to the reduced finite-sample
efficiency. In this case, the empirical standard deviation (0.0459) exceeded the average estimated standard error (0.0420), although coverage remained close to nominal level at 94.6\%. The proposed estimator generally approached the variability of the oracle benchmark as the sample size increased.

For prioritized semi-competing risks, the proposed estimator had relative efficiency between 1.84 and 2.06 across all scenarios, representing a reduction of approximately 46\%--51\% in empirical variance compared with IPCW. Relative efficiency was also somewhat smaller at $n=200$ (1.84--1.85) than at larger sample sizes,
although substantial efficiency gains were retained. This pattern in both outcome settings is consistent with
greater finite-sample variability in estimation of the nuisance functions within the cross-fitting training samples, which may limit the realized efficiency gains at smaller sample sizes. The empirical standard deviations of the proposed estimator were close to those of the oracle benchmark, particularly at larger sample sizes, and the  standard errors produced coverage between 94.5\% and 95.3\%. These findings indicate that incorporating the prognostic
baseline covariate and observed event history can substantially improve precision under the simulated censoring mechanisms.

\begin{table}[htbp]
\centering
\scriptsize
\begin{threeparttable}
\caption{Finite-sample performance for marginal win probability in the scalar right-censored setting.}
\label{tab:sim_scalar}
\setlength{\tabcolsep}{4pt}
\begin{tabular}{lllrrrrrr}
\toprule
Scenario & $n$ & Estimator & True $\psi$ & Bias & ESD & ASE & Coverage & RE \\
\midrule
Moderate & 200 & IPCW   & 0.5548 &  0.0000 & 0.0405 & 0.0410 & 0.952 & 1.00 \\
         &     & EIF    &        & -0.0001 & 0.0377 & 0.0380 & 0.953 & 1.15 \\
         &     & Oracle &        & -0.0001 & 0.0367 & 0.0371 & 0.952 & 1.21 \\
\addlinespace[0.25em]
         & 400 & IPCW   & 0.5548 & -0.0001 & 0.0285 & 0.0287 & 0.950 & 1.00 \\
         &     & EIF    &        &  0.0001 & 0.0264 & 0.0265 & 0.950 & 1.17 \\
         &     & Oracle &        &  0.0002 & 0.0261 & 0.0262 & 0.952 & 1.19 \\
\addlinespace[0.25em]
         & 800 & IPCW   & 0.5548 &  0.0001 & 0.0202 & 0.0203 & 0.951 & 1.00 \\
         &     & EIF    &        &  0.0001 & 0.0185 & 0.0186 & 0.956 & 1.19 \\
         &     & Oracle &        &  0.0001 & 0.0185 & 0.0185 & 0.955 & 1.20 \\
\midrule
Stronger & 200 & IPCW   & 0.5529 &  0.0004 & 0.0476 & 0.0476 & 0.946 & 1.00 \\
         &     & EIF    &        &  0.0006 & 0.0459 & 0.0420 & 0.946 & 1.08 \\
         &     & Oracle &        &  0.0004 & 0.0393 & 0.0394 & 0.950 & 1.47 \\
\addlinespace[0.25em]
         & 400 & IPCW   & 0.5529 & -0.0005 & 0.0330 & 0.0333 & 0.954 & 1.00 \\
         &     & EIF    &        &  0.0000 & 0.0286 & 0.0287 & 0.951 & 1.33 \\
         &     & Oracle &        &  0.0001 & 0.0277 & 0.0278 & 0.953 & 1.41 \\
\addlinespace[0.25em]
         & 800 & IPCW   & 0.5529 &  0.0001 & 0.0228 & 0.0233 & 0.952 & 1.00 \\
         &     & EIF    &        &  0.0000 & 0.0196 & 0.0199 & 0.951 & 1.35 \\
         &     & Oracle &        &  0.0000 & 0.0194 & 0.0196 & 0.953 & 1.37 \\
\midrule
Strongest & 200 & IPCW   & 0.5432 &  0.0005 & 0.0465 & 0.0467 & 0.949 & 1.00 \\
          &     & EIF    &        &  0.0004 & 0.0341 & 0.0336 & 0.946 & 1.85 \\
          &     & Oracle &        &  0.0004 & 0.0328 & 0.0326 & 0.949 & 2.01 \\
\addlinespace[0.25em]
          & 400 & IPCW   & 0.5432 &  0.0000 & 0.0328 & 0.0328 & 0.950 & 1.00 \\
          &     & EIF    &        &  0.0003 & 0.0232 & 0.0233 & 0.954 & 1.99 \\
          &     & Oracle &        &  0.0004 & 0.0229 & 0.0229 & 0.952 & 2.05 \\
\addlinespace[0.25em]
          & 800 & IPCW   & 0.5432 & -0.0002 & 0.0228 & 0.0231 & 0.953 & 1.00 \\
          &     & EIF    &        & -0.0001 & 0.0159 & 0.0163 & 0.956 & 2.04 \\
          &     & Oracle &        & -0.0001 & 0.0159 & 0.0162 & 0.954 & 2.05 \\
\bottomrule
\end{tabular}
\begin{tablenotes}[flushleft]
\scriptsize
\item[] EIF: the proposed estimator based on efficient influence function;
Oracle: the estimator using the true event and censoring hazards while
estimating the remaining nuisance quantities;
ESD: empirical standard deviation;
ASE: average estimated standard error;
Coverage: empirical coverage of nominal 95\% confidence intervals;
RE: empirical variance of IPCW divided by that of the indicated estimator.
\end{tablenotes}
\end{threeparttable}
\end{table}

\begin{table}[htbp]
\centering
\scriptsize
\begin{threeparttable}
\caption{Finite-sample performance for marginal win probability in the semi-competing risk setting.}
\label{tab:sim_practical}
\setlength{\tabcolsep}{4pt}
\begin{tabular}{lllrrrrrr}
\toprule
Scenario & $n$ & Estimator & True $\psi$ & Bias & ESD & ASE & Coverage & RE \\
\midrule
Moderate & 200 & IPCW   & 0.5554 & -0.0003 & 0.0433 & 0.0439 & 0.947 & 1.00 \\
         &     & EIF    &        & -0.0001 & 0.0318 & 0.0322 & 0.950 & 1.85 \\
         &     & Oracle &        &  0.0002 & 0.0312 & 0.0317 & 0.954 & 1.92 \\
\addlinespace[0.25em]
         & 400 & IPCW   & 0.5554 &  0.0002 & 0.0313 & 0.0306 & 0.944 & 1.00 \\
         &     & EIF    &        &  0.0008 & 0.0228 & 0.0223 & 0.947 & 1.89 \\
         &     & Oracle &        &  0.0008 & 0.0226 & 0.0222 & 0.946 & 1.91 \\
\addlinespace[0.25em]
         & 800 & IPCW   & 0.5554 &  0.0003 & 0.0214 & 0.0215 & 0.946 & 1.00 \\
         &     & EIF    &        &  0.0003 & 0.0156 & 0.0157 & 0.953 & 1.89 \\
         &     & Oracle &        &  0.0003 & 0.0155 & 0.0156 & 0.952 & 1.90 \\
\midrule
Stronger & 200 & IPCW   & 0.5557 & -0.0004 & 0.0455 & 0.0463 & 0.952 & 1.00 \\
         &     & EIF    &        &  0.0000 & 0.0335 & 0.0340 & 0.952 & 1.84 \\
         &     & Oracle &        &  0.0003 & 0.0329 & 0.0332 & 0.950 & 1.91 \\
\addlinespace[0.25em]
         & 400 & IPCW   & 0.5557 &  0.0004 & 0.0326 & 0.0322 & 0.950 & 1.00 \\
         &     & EIF    &        &  0.0010 & 0.0237 & 0.0235 & 0.945 & 1.88 \\
         &     & Oracle &        &  0.0010 & 0.0237 & 0.0233 & 0.945 & 1.90 \\
\addlinespace[0.25em]
         & 800 & IPCW   & 0.5557 &  0.0005 & 0.0227 & 0.0227 & 0.949 & 1.00 \\
         &     & EIF    &        &  0.0005 & 0.0164 & 0.0164 & 0.951 & 1.93 \\
         &     & Oracle &        &  0.0005 & 0.0163 & 0.0164 & 0.952 & 1.94 \\
\midrule
Strongest & 200 & IPCW   & 0.5558 & -0.0010 & 0.0500 & 0.0521 & 0.956 & 1.00 \\
          &     & EIF    &        & -0.0001 & 0.0362 & 0.0367 & 0.952 & 1.91 \\
          &     & Oracle &        &  0.0001 & 0.0352 & 0.0353 & 0.952 & 2.02 \\
\addlinespace[0.25em]
          & 400 & IPCW   & 0.5558 &  0.0002 & 0.0357 & 0.0357 & 0.949 & 1.00 \\
          &     & EIF    &        &  0.0009 & 0.0253 & 0.0251 & 0.947 & 1.99 \\
          &     & Oracle &        &  0.0009 & 0.0251 & 0.0247 & 0.947 & 2.03 \\
\addlinespace[0.25em]
          & 800 & IPCW   & 0.5558 &  0.0002 & 0.0250 & 0.0249 & 0.949 & 1.00 \\
          &     & EIF    &        &  0.0004 & 0.0174 & 0.0175 & 0.950 & 2.06 \\
          &     & Oracle &        &  0.0004 & 0.0173 & 0.0174 & 0.952 & 2.08 \\
\bottomrule
\end{tabular}
\begin{tablenotes}[flushleft]
\scriptsize
\item[] EIF: the proposed estimator based on efficient influence function;
Oracle: the estimator using the true event and censoring hazards while
estimating the remaining nuisance quantities;
ESD: empirical standard deviation;
ASE: average estimated standard error;
Coverage: empirical coverage of nominal 95\% confidence intervals;
RE: empirical variance of IPCW divided by that of the indicated estimator.
\end{tablenotes}
\end{threeparttable}
\end{table}

\section{Real Data Application}

We applied the proposed method to breast cancer clinical trial data described by \citet{Collett2023} and made available
through the R package \texttt{collett}. This was a randomized clinical trial of breast cancer patients comparing tamoxifen combined with radiotherapy against tamoxifen alone. The data are publicly available
in the \texttt{tamoxifen} dataset of the R package \texttt{collett}. The trial recorded overall survival and several types of disease relapse, including local, axillary, and distant relapse, providing
an opportunity to evaluate treatment effects using a prioritized outcome in the presence of censoring.

For the present analysis, treatment was coded as \(A=1\) for tamoxifen plus radiotherapy and \(A=0\) for tamoxifen alone.
We considered death as the terminal event with higher priority and the first occurrence of local, axillary, or distant relapse before death as the nonterminal event with lower priority. When relapse and death occurred on the same day, death was given priority. We examined restricted outcomes at horizons of \(\tau=4\) and \(5\) years. The estimand was the marginal probability that a randomly selected patient assigned to tamoxifen plus radiotherapy had a more favorable restricted prioritized outcome than a randomly selected patient assigned to tamoxifen alone, with half credit assigned to ties.

Baseline adjustment included age at study entry, tumor size, and hormone-receptor status. Age was centered at 60 years and scaled by 10, and tumor size was transformed using $\log(1+\text{size})$. These covariates were included in the
working models for the event and censoring processes. The death and relapse hazards were modeled separately by
treatment arm, with time-varying baseline hazards represented using penalized cubic B-splines. The death hazard additionally depended on the time-dependent indicator $N_2(t)=I(T_2<t)$, allowing the death hazard to change
following relapse. The censoring hazard was modeled using a pooled model including treatment, baseline covariates,
and $N_2(t)$. The baseline log hazards were approximated using 16 time intervals and cubic B-splines with one interior
knot at the midpoint of follow-up. Ridge regularization with penalty parameter $\kappa=2$ was applied to the
regression coefficients, and a second-difference penalty with parameter $\lambda=8$ was applied to the spline coefficients.

We compared the IPCW estimator with the proposed cross-fitted one-step estimator. Five-fold cross-fitting was used to estimate the nuisance functions. The treatment assignment probability was taken to be $1/2$, consistent with the randomized design. The functions $h_1$ and $h_0$ were constructed using IPCW estimates of the marginal distributions of the restricted prioritized outcomes. The conditional means $\mu_1$ and $\mu_0$, together with the residual projections
$q_a(u,\mathcal H_u)$, were estimated using the fitted event-history models. In particular, the full censoring
augmentation involving $q_a$ was retained and evaluated using deterministic numerical integration. Standard errors
for the proposed estimator were obtained  from the estimated efficient influence function. For the IPCW estimator, standard errors were computed using a first-order influence-function approximation
accounting for estimation of the penalized censoring model.

The analysis included 641 patients, with 321 assigned to tamoxifen alone and 320 assigned to tamoxifen plus radiotherapy.
At the four-year horizon, 17.2\% of patients were censored before their restricted prioritized outcomes were fully
observed. The IPCW estimator gave a win probability of $0.5270$, corresponding to win odds of $1.1140$, with
a standard error of $0.0138$ for the win probability and a 95\% confidence interval of $(0.9997,1.2412)$
for the win odds. The proposed one-step estimator gave a win probability of $0.5242$, corresponding to win odds
of $1.1016$, with a standard error of $0.0125$ and a 95\% confidence interval of $(0.9987,1.2150)$ for
the win odds. The corresponding two-sided $p$ values for testing a log win odds of zero were $0.0506$ and
$0.0532$, respectively. The ratio of squared estimated standard errors for IPCW relative to the one-step estimator
was $1.22$.

At the five-year horizon, the proportion of censored patients increased to 34.0\%. The IPCW estimator gave a
win probability of $0.5365$, corresponding to win odds of $1.1574$, with a standard error of $0.0154$
and a 95\% confidence interval of $(1.0254,1.3063)$ for the win odds. The proposed one-step estimator gave a
win probability of $0.5300$, which gave a win odds of $1.1276$, with a standard error of $0.0133$
and a 95\% confidence interval of $(1.0159,1.2517)$ for the win odds. The corresponding two-sided $p$ values
were $0.0180$ and $0.0241$, respectively. The estimated variance ratio comparing IPCW with the one-step estimator
was $1.34$, corresponding to an approximately 25\% reduction in estimated variance.

Based on these results, both estimators favored tamoxifen plus radiotherapy with respect to the prioritized outcomes, although the estimated treatment effects were relatively small. The treatment comparison was statistically significant at
five years but not at four years at the 5\% significance level. The proposed estimator yielded smaller estimated
standard errors at both horizons, with a more pronounced improvement at five years. These findings illustrate the
potential efficiency gain from incorporating baseline prognostic information and the censoring augmentation in
the analysis of prioritized outcomes. The estimated variance ratios describe the relative precision observed in this
dataset, and their interpretation depends on the adequacy of the working nuisance models and the assumptions underlying
the influence-function-based inference.

We conducted sensitivity analyses to assess the stability of the results with respect to the spline specifications.
At the five-year horizon, we varied the smoothing penalty parameter from $\lambda=2$ to $\lambda=20$ while retaining
one interior knot, and also considered a more flexible specification with two interior knots and $\lambda=8$.
Across the four specifications, the estimated win probabilities ranged from $0.5299$ to $0.5301$, with
standard errors ranging from $0.01325$ to $0.01327$. The estimated variance ratios comparing IPCW with the
proposed estimator remained approximately $1.34$. The minimum estimated censoring survival probabilities
were also stable across specifications. These results suggest that the estimated treatment effect
and efficiency improvement are relatively insensitive to the considered spline specifications.

As an additional sensitivity analysis, we fitted exponential working models for the event and censoring hazards using the same estimation procedure. At the four-year time horizon, the proposed one-step estimator yielded a win
probability of $0.5230$ with a standard error of $0.0129$, compared with $0.5242$ and $0.0125$ under the spline models.
At the five-year horizon, the corresponding estimates were $0.5306$ and $0.0144$, compared with $0.5300$ and $0.0133$
under the spline models. The estimated variance ratios comparing IPCW with the proposed estimator were $1.19$
and $1.34$ at four and five years, respectively. These results indicate that the estimated treatment effects
and efficiency gains are relatively insensitive to the choice between the parametric (exponential) and more flexible spline baseline hazards models.

\section{Discussion}

We developed a semiparametric framework for covariate-adjusted win-type analysis with censored time-to-event outcomes. The proposed estimand is a marginal win probability comparing restricted outcomes between treatment arms, with the
corresponding win odds providing a summary on the odds scale. We derived the  efficient influence function of the win probability under conditional independent censoring given the observed history, and proposed a cross-fitted one-step estimator that incorporates both baseline covariate adjustment and a censoring-martingale augmentation. The proposed estimator allows flexible estimation of nuisance functions, and its standard errors can be obtained directly from the estimated influence function without bootstrap resampling. The
simulations demonstrated small bias and accurate estimated standard errors and coverage of confidence intervals. The proposed estimator was consistently more efficient than IPCW, with particularly substantial
efficiency gains in the semi-competing risk setting.

The proposed framework focuses on the marginal win probability and the corresponding win odds, which is motivated by the causal interpretation of the estimand. The estimand is defined by comparing two independent marginal
potential outcomes, one drawn from the treatment distribution and one drawn
from the control distribution, rather than by comparing the two potential outcomes
from the same subject. This distinction is important because, even in a randomized trial, the joint
distribution of the two potential outcomes for the same individual is not
identified from the observed data without additional assumptions. In contrast, the marginal comparison between
independent draws from the distributions of the two treatment arms  is identifiable
from the randomized trial and has a direct probabilistic-index interpretation. This marginal interpretation is consistent with the formulation of win-type estimands discussed in \cite{Mao2018}.
The same general framework could be extended to the win ratio. Such an extension would require defining the marginal win and loss probabilities separately and deriving the influence function for their ratio. Although this
extension is conceptually straightforward, it would add notational and technical complexity. We therefore focus on the win probability and win odds in the present work.

As with other inverse-probability-weighting methods, the proposed estimator may be sensitive to very small censoring
survival probabilities, which can produce extreme inverse-censoring weights and unstable contributions to the influence functions, even when the theoretical positivity assumption is satisfied. In practice, we recommend examining
the distribution of estimated censoring probabilities and weights and assessing the sensitivity of the results to
the specification of the models for the nuisance functions. Regularization can improve numerical stability, as illustrated by the ridge-penalized hazard estimation used in our numerical
studies. Although weight stabilization or truncation may also mitigate instability, truncation can introduce bias
and may affect the validity of the censoring-martingale augmentation. We therefore did not impose weight truncation
in our simulations or real data analysis. Furthermore, the asymptotic guarantees depend on the consistency and
convergence rates of the estimators nuisance parameters. These conditions should be considered when applying the method
with highly flexible nuisance estimation procedures.

\section*{Code and Data Availability}

An R package, \texttt{winoddsEIF}, implementing the proposed method, together with code for reproducing the simulation studies and real-data analysis, is publicly available at \url{https://github.com/zhiguoli919/winoddsEIF}. The repository includes documentation and instructions for reproducing the reported results. The breast cancer trial data used in the application are available in the R package \texttt{collett}.


\begin{thebibliography}{99}

\bibitem[Bebu and Lachin(2016)]{BebuLachin2016}
Bebu, I. and Lachin, J. M. (2016).
Large sample inference for a win ratio analysis of a composite outcome
based on prioritized components.
\textit{Biostatistics}, 17, 178--187.


\bibitem[Benkeser et al.(2021)]{Benkeser2021}
Benkeser, D., D\'iaz, I., Luedtke, A., Segal, J., Scharfstein, D.,
and Rosenblum, M. (2021).
Improving precision and power in randomized trials for COVID-19 treatments
using covariate adjustment, for binary, ordinal, and time-to-event outcomes.
\textit{Biometrics}, 77, 1467--1481.

\bibitem[Bickel et al.(1993)]{Bickel1993}
Bickel, P. J., Klaassen, C. A. J., Ritov, Y., and Wellner, J. A. (1993).
\textit{Efficient and Adaptive Estimation for Semiparametric Models}.
Johns Hopkins University Press, Baltimore.

\bibitem[Brunner et al.(2021)]{Brunner2021}
Brunner, E., Vandemeulebroecke, M., and M\"utze, T. (2021).
Win odds: an adaptation of the win ratio to include ties.
\textit{Statistics in Medicine}, 40, 3367--3384.


\bibitem[Buyse(2010)]{Buyse2010}
Buyse, M. (2010).
Generalized pairwise comparisons of prioritized outcomes in the two-sample problem.
\textit{Statistics in Medicine}, 29, 3245--3257.


\bibitem[Cao et al.(2025)]{Cao2025}
Cao, Z., Zuo, S., Baumann, M. R., Plourde, K., Heagerty, P.,
Tong, G., and Li, F. (2025).
Covariate-adjusted win statistics in randomized clinical trials with ordinal outcomes.
\textit{arXiv preprint arXiv:2508.20349}.

\bibitem[Chernozhukov et al.(2018)]{Chernozhukov2018}
Chernozhukov, V., Chetverikov, D., Demirer, M., Duflo, E.,
Hansen, C., Newey, W., and Robins, J. (2018).
Double/debiased machine learning for treatment and structural parameters.
\textit{The Econometrics Journal}, 21, C1--C68.

\bibitem[Collett(2023)]{Collett2023}
Collett, D. (2023).
\textit{Modelling Survival Data in Medical Research},
4th ed.
Chapman \& Hall/CRC, Boca Raton, FL.

\bibitem[Dandl and Hothorn(2026)]{DandlHothorn2026}
Dandl, S. and Hothorn, T. (2026).
Nonparanormal adjusted marginal inference.
\textit{Biometrics}, 82, ujag065.

\bibitem[Dong et al.(2020)]{Dong2020IPCW}
Dong, G., Mao, L., Huang, B., Gamalo-Siebers, M., Wang, J.,
Yu, G., and Hoaglin, D. C. (2020).
The inverse-probability-of-censoring weighting (IPCW) adjusted win ratio
statistic: an unbiased estimator in the presence of independent censoring.
\textit{Journal of Biopharmaceutical Statistics}, 30, 882--899.

\bibitem[Dong et al.(2021)]{Dong2021Dependent}
Dong, G., Huang, B., Wang, D., Verbeeck, J., Wang, J.,
and Hoaglin, D. C. (2021).
Adjusting win statistics for dependent censoring.
\textit{Pharmaceutical Statistics}, 20, 440--450.

\bibitem[Fleming and Harrington(1991)]{FlemingHarrington1991}
Fleming, T. R. and Harrington, D. P. (1991).
\textit{Counting Processes and Survival Analysis}.
Wiley, New York.


\bibitem[FDA(2023)]{FDA2023}
U.S. Food and Drug Administration. (2023).
\textit{Adjusting for Covariates in Randomized Clinical Trials for Drugs
and Biological Products: Guidance for Industry}.
U.S. Food and Drug Administration.

\bibitem[Huang et al.(2026)]{Huang2026}
Huang, Z., Shao, L., Guo, Y., Lu, Y., Tu, X. M.,
Feng, C., and Lin, T. (2026).
Win-ratio regression for prioritized composite outcomes
in observational studies: Doubly robust and efficient
estimation with future-score correction.
\textit{arXiv preprint arXiv:2608.10728}, version 2.

\bibitem[Koziol and Jia(2009)]{KoziolJia2009}
Koziol, J. A. and Jia, Z. (2009).
The concordance index \(C\) and the Mann--Whitney parameter
\(\Pr(X>Y)\) with randomly censored data.
\textit{Biometrical Journal}, 51, 467--474.


\bibitem[Mao(2018)]{Mao2018}
Mao, L. (2018).
On causal estimation using U-statistics.
\textit{Biometrika}, 105, 215--220.

\bibitem[Martinussen and Vansteelandt(2025)]{Martinussen2025}
Martinussen, T. and Vansteelandt, S. (2025).
Debiased learning of the causal net benefit with censored
event time data.
\textit{Biometrika}, 112, asaf051.

\bibitem[Pocock et al.(2012)]{Pocock2012}
Pocock, S. J., Ariti, C. A., Collier, T. J., and Wang, D. (2012).
The win ratio: a new approach to the analysis of composite endpoints in
clinical trials based on clinical priorities.
\textit{European Heart Journal}, 33, 176--182.

\bibitem[Pocock et al.(2024)]{Pocock2024}
Pocock, S. J., et al. (2024).
The win ratio in cardiology trials: lessons learnt, new developments,
and wise future use.
\textit{European Heart Journal}, 45, 4684--4697.


\bibitem[Scheidegger et al.(2026)]{Scheidegger2026}
Scheidegger, C., Wandel, S., and M\"utze, T. (2026).
Covariate adjustment for the win odds: application to cardiovascular
outcomes trials.
\textit{Statistics in Medicine}, 45, e70602.

\bibitem[Tsiatis(2006)]{Tsiatis2006}
Tsiatis, A. A. (2006).
\textit{Semiparametric Theory and Missing Data}.
Springer, New York.


\bibitem[Tsiatis et al.(2008)]{Tsiatis2008}
Tsiatis, A. A., Davidian, M., Zhang, M., and Lu, X. (2008).
Covariate adjustment for two-sample treatment comparisons in randomized
clinical trials: a principled yet flexible approach.
\textit{Statistics in Medicine}, 27, 4658--4677.


\bibitem[van der Vaart(1998)]{vanderVaart1998}
van der Vaart, A. W. (1998).
\textit{Asymptotic Statistics}.
Cambridge University Press, Cambridge.

\bibitem[van der Laan and Robins(2003)]{vanderLaanRobins2003}
van der Laan, M. J. and Robins, J. M. (2003).
\textit{Unified Methods for Censored Longitudinal Data and Causality}.
Springer, New York.


\bibitem[Zhang et al.(2008)]{Zhang2008}
Zhang, M., Tsiatis, A. A., and Davidian, M. (2008).
Improving efficiency of inferences in randomized clinical trials using
auxiliary covariates.
\textit{Biometrics}, 64, 707--715.

\end{thebibliography}
\end{document}